\documentclass{emulateapj}
\shorttitle{Brightness suppression of quasar radio jets}
\shortauthors{Gopal-Krishna et al.}
\received{2003 December 29; accepted, ApJ Letters 2004 January 21}
\begin{document}
\title{Brightness suppression of relativistic radio jets of quasars:\\
The role of the lower electron energy cut-off }

\author{Gopal-Krishna\altaffilmark{1}}
%\and  %COMMENT OUT THIS "\AND" FOR EMULATEAPJ
\author{Peter L.\ Biermann} 
\affil{Max-Planck-Institut f{\"u}r Radioastronomie, Auf dem H{\"u}gel 69, 
D-53121 Bonn, Germany}
\email{krishna@ncra.tifr.res.in}

%\author{P.\ L.\ Biermann} 
%\affil{Max-Planck-Institut f{\"u}r Radioastronomie, Auf dem H{\"u}gel 69, 
%53121 Bonn, Germany}
\email{plbiermann@mpifr-bonn.mpg.de}
\and
\author{Paul J.\ Wiita\altaffilmark{2}}
\affil{Department of Physics \& Astronomy, MSC 8R0314,
Georgia State University, 33 Gilmer Street SE, Unit 8,
Atlanta, GA 30303-3088}
\email{wiita@chara.gsu.edu}

%\offprints{Gopal-Krishna
%            \email{krishna@ncra.tifr.res.in}}
\altaffiltext{1}{Alexander von Humboldt Senior Fellow; on leave from the National
Centre for Radio Astrophysics/TIFR,
Pune, India}
\altaffiltext{2}{also, Department of Astrophysical Sciences, Princeton
University}

\begin{abstract}
Although the possibility of a low-energy cut-off (LEC) of the relativistic 
electron population (E$_{min} \sim 0.1$ GeV) in the different components of 
radio galaxies has been discussed in the literature, from both theoretical 
and observational viewpoints, its possible role in causing a distinct 
spectral flattening, and thus reducing the apparent brightness temperature 
of the radio nucleus, has thus far not been explicitly discussed. Here we 
point out that such an effect may in fact be quite significant for the 
parsec-scale, superluminal emission regions associated with the inner radio jets. 
This is because the observed frequency of the  spectral turnover arising from the 
LEC increases linearly with the bulk Doppler factor ($\delta$) of the 
jet flow, whereas the frequency of synchrotron self-absorption (the 
well-known cause of spectral flattening and turnover at low frequencies) 
shows a mild decrease as 
$\delta$ rises. For an observationally 
relevant range of source parameters, we investigate here the role of the 
LEC.  We argue that some statistical trends revealed by the 
recent very long baseline interferometry (VLBI) surveys can in fact be understood in terms of effects arising
from LEC, including the faster superluminal motion found in the VLBI surveys 
at higher frequencies, and the apparent inability of most radio 
cores to even attain the brightness temperatures consistent with the 
equipartition
condition.  We also point out some possible implications of the 
LEC for X-ray observations of %the kiloparsec scale relativistic jets of quasars
%made with {\sl Chandra} and {\sl XMM-Newton}.
kiloparsec scale relativistic jets of quasars.

\end{abstract}


\noindent Submitted 29 December 2003; accepted by
Astrophysical Journal Letters on 21 January 2004

\section{Introduction}

The frequent association of flat or inverted radio spectra with the 
parsec-scale knots in the jets of luminous active galactic nuclei
(AGN; BL Lacertae objects
and quasars) is 
commonly interpreted in terms of synchrotron self-absorption (SSA) effects. 
A combination of  very long baseline interferometry (VLBI) 
measurements of the angular size and flux density 
near the frequency of SSA, with the X-ray flux density expected from 
synchrotron self-Compton scattering, then yields the values of the 
magnetic field, $B$, and the Doppler factor, $\delta = 1/[\Gamma - {\rm cos} 
{\phi}  ({\Gamma^2} - 1)^{1/2}]$, for the knots, where $\Gamma$ is the bulk 
Lorentz factor and $\phi$ is the viewing angle to the axis of the flow; 
$\delta$ has a maximum value equal to 2$\Gamma$ (e.g., Marscher 1987; 
Ghisellini et al.\ 1992).  Evidence for spectral flattening at lower 
radio frequencies, possibly attributable to free-free absorption in the nuclear 
region, has also been discussed in the context of a few superluminal AGN 
(e.g., Matveyenko et al.\ 1980; Britzen et al.\ 2001; %Bicknell xxxx;
Kellermann 2003, and references therein).  

Here we point out that a significant spectral flattening of the superluminal 
radio knots could also arise from a low-energy cut-off (LEC) of the 
relativistic electron population and, within a plausible range of 
parameters for some VLBI components, this can mimic SSA even in  
spectral regions that are essentially free from SSA.
The resulting extra attenuations of the radio flux and apparent brightness 
temperature may have interesting implications, and seem to provide useful 
clues to some observational trends revealed by recent VLBI surveys, 
which are briefly pointed out below.

%\section {Low Energy Cut-offs vs.\ Synchrotron Self-Absorption}
\section {LEC vs.\ SSA}

From the basic theory of synchrotron emission it is well known that an 
energy distribution of relativistic electrons truncated at low energies 
gives rise to a  spectral turnover toward lower frequencies, 
attaining a spectral index $\alpha = +1/3$ (defined so that 
the flux density, $S_{\nu} \propto \nu^{\alpha}$) (e.g. Ginzburg \& 
Syrovatskii 1969; Pacholczyk 1970). %; Salter \& Brown 1988).  
In the rest 
frame of the emitting region this cutoff frequency, $\nu_m = 0.29 \nu_c$, 
where $\nu_c$(Hz) = $6.27 \times 10^{18}~ B ~{\rm sin}\Theta ~E_{min}^2$, 
where $B$ is in units of gauss, $E_{min} = \gamma_{min}m_e c^2$, in ergs,
is the LEC for the electron distribution, 
and  $\Theta$ is the mean pitch angle of the 
electrons with respect to the magnetic field.  If we assume a
nearly isotropic 
distribution of electrons in the rest frame of the emitting region, we 
find that the spectrum turns over below an observed frequency 
\begin{equation}
\nu_m ({\rm GHz}) \simeq 30 B \Bigl({\frac{E_{min}}{100 MeV}}\Bigr)^2
\Bigl({\frac{\delta}{1 + z}}\Bigr), 
\end{equation}
with  $z$  the cosmological redshift of the AGN. 

A few rare examples for which hints of such a low-frequency spectral
turnover have been reported include the hot spots of the bright radio galaxy 
Cyg A, from which $\gamma_{min} = 350 \pm 50$ has  been inferred 
(Biermann, Strom \& Falcke 1995).  Similar values ($\gamma_{min} \sim 500-
1000$) have been inferred by \cite{cari} for Cyg A and by
\cite{hard} for 3C 196;  Leahy, Muxlow \& Stephens
(1989) give indications of a LEC in other  hot spots. 
However, %even in this `leading' case, 
hot spot evidence for a LEC 
remains weak, mainly because one really needs flux measurements with 
arcsecond resolution well below the turnover frequency $\nu_{SSA} \simeq$ 
200 MHz.  Given the practical difficulty of obtaining detailed images at 
such low frequencies, we explore here the possibility of investigating
this phenomenon using the radio knots in AGN jets whose apparent 
superluminal motion can push the spectral turnover peak due to LEC 
(near $\nu_m$) 
into the readily accessible centimeter/millimeter waveband. It may thus 
become much more feasible to trace the observational signatures of LEC.
Clearly, any such evidence would hold vital clues to the particle 
acceleration mechanism in AGN and also shed some light on other basic 
questions concerning the energetics of the AGN jets.

Probably the most secure example of a mildly inverted radio spectrum
arising from LEC is the nuclear source of our own Galaxy, Sgr A$^*$,
whose radio spectrum has a slope close to $\alpha = +1/3$. This has been 
interpreted in terms of a quasi mono-energetic distribution of the radiating 
electron population, peaked near 10$^2$ MeV (Duschl \& Lesch 1994; Beckert 
\& Duschl 1997;  Falcke, Mannheim \& Biermann 1993). 
One specific proposal that yields such a LEC
invokes hadronic interactions near the inner edge of the accretion
disk around a rapidly spinning central black hole
where it connects to the jet (Falcke, Malkan \& Biermann 1995;
Donea \& Biermann 1996).  This mechanism produces
a pair plasma embedded in a normal plasma; the LEC would 
correspond to the pion mass, at least near the zone of creation,
subject to adiabatic losses and reacceleration along the jet.

In the context of blazars, an early hint for LEC (with $E_{min}
\approx 100$ 
MeV) came from the observed lack of Faraday depolarization in the VLBI emission
regions (Wardle 1977; Jones \& O'Dell 1977). Interestingly, this value of $E_{min}$ 
is also the typical energy of electrons radiating near the self-absorption 
frequency (e.g., Celotti \& Fabian 1993).  \cite{ghis} argue 
that synchrotron emission in VLBI scale jets may well arise from $e^{\pm}$ 
pairs and obtain a good fit, taking a LEC of around 50 MeV.  
\cite{celo93} conversely argue that energetics constraints
indicate that radio jets are `heavy' and are composed of protons and 
electrons, but they still need to invoke a LEC.  Each of these arguments 
is model-dependent, and the value of $E_{min}$ 
%the cut-off energy 
is yet to be firmly 
established. Likewise, the mechanism responsible for a LEC remains 
to be understood.  It was argued long ago \citep{jone} that the 
existence of a LEC favors models in which radiating relativistic particles 
are accelerated in relativistic blasts (e.g., Blandford \& McKee 1976) or 
by strong electromagnetic acceleration (e.g., Lovelace 1976), rather than 
through stochastic acceleration, since the latter mechanism should produce 
many electrons with only modestly relativistic Lorentz factors. But neither 
of these approaches naturally produces a  LEC around 50--100 MeV. 
%(or $100 \la \gamma_{min} \la 200)$. 
The arguments presented here are quite independent of the 
specific mechanism for a LEC.

%One specific proposal that yields a LEC
%invokes hadronic interactions near the inner edge of the accretion
%disk where it connects to the jet (Falcke, Malkan \& Biermann 1995;
%Donea \& Biermann 1996).  Transferring energy from the disk
%to the jet turns the inner portion of the disk into an advection
%dominated/radiatively inefficient accretion flow, whose temperature
%can rise enough to cause hadronic interactions \citep{maha}.
%The temperatures become relativistic if the black hole is
%close to maximally rotating, so in this specific scenario, the 
%LEC is an indicator of the black hole's spin.  This argument
%has guided recent work on the angular momentum transport by
%gravitational waves during the merger of two black holes
%(Chirvasa 2002; Biermann et al.\ 2003).  This mechanism yields
%a pair plasma embedded in a normal plasma, and the LEC would 
%correspond to the pion mass, at least near the zone of creation,
%subject to adiabatic losses and reacceleration along the jet.
%Even though $E_{min} \approx$ 100 MeV is a natural feature of this last 
%scenario, 
%we note that the arguments presented here are quite independent of the 
%specific mechanism for a LEC. 

For a homogeneous, spherical source of incoherent synchrotron emission 
SSA leads to a rather sharp spectral turnover 
near a frequency (e.g., Ginzburg \& Syrovatskii 1969; Kellermann \& 
Pauliny-Toth 1969)  
\begin{equation}
\nu_{SSA} ({\rm GHz}) = 8 \Bigl({\frac{B}{\delta_0}}\Bigr)^{1/5}
\Bigl({\frac{S_{max}}{\theta^2}}\Bigr)^{2/5},
\end{equation}
where $S_{max}$ is the peak flux density
(in units of janskys) of the knot, observed at $\nu_{SSA}$, $\theta$ 
(in milli-arcseconds) 
is its angular size at $\nu_{SSA}$, $B$ is in units of gauss, and $\delta_0 = \delta/(1+z)$ is the 
co-moving Doppler factor of the knot.

From Eq (1) and Eq (2) we get
\begin{equation}
\eta \equiv \nu_m / \nu_{SSA} = 4 ~\delta_0^{6/5} ~ B^{4/5} ~E_{min,2}^2 ~\Sigma^{-2/5}.
\end{equation}
Here $\Sigma \equiv S_{max}/\theta^2 $ is a measure of the 
apparent surface brightness, and $E_{min,2} \equiv E_{min}/100{\rm MeV}$. 
The critical Doppler factor for which $\eta = 1$ is given by
\begin{equation}
\delta_{\star} = 0.3~ (1 + z)~  B^{-2/3}~ E_{min,2}^{-5/3}~ \Sigma^{1/3};   
\end{equation}
if $\delta \gtrsim \delta_{\star}$ then LEC is important for flux 
attenuation.  Although our picture of the knots is admittedly
oversimplified, as any gradients in the LEC could complicate the analysis,
it should suffice to illustrate the relevance
of LEC induced spectral effects.  Clearly, for the 
knots with $\eta > 1$, the onset of (a mild) spectral turnover toward lower 
frequencies will be led by LEC, not SSA. 
%(instead of the SSA, which would be effective only at lower frequencies). 
Thus, the larger the value of $\eta$, the greater 
would be the suppression of the peak flux that would have otherwise been 
attained by the knot, provided the limiting factor were indeed SSA. Considering the 
mildly positive slope ($\alpha \simeq 1/3$) at $\nu < \nu_m$ and a typical 
spectral index $\alpha \simeq -0.7$ in the unabsorbed/uncutoff part of the spectrum, 
the flux suppression factor measured at $\nu_{SSA}$ is
$\psi \simeq \eta^{1/3-\alpha} \approx \eta$.
%, if one defines that suppression factor %ratio
%as measured at $\nu_{SSA}$.
%, or $\psi^{\prime} \simeq \eta^{-\alpha}$,
%if one defines it as the cutoff maximal flux at $\nu_m$, when compared 
%to what the peak flux would have been at $\nu_{SSA}$ in the absence of a LEC.

The source parameters on which $\eta$ and $\psi$ depend include the magnetic 
field in 
the VLBI knot, for which estimates are rather model dependent. For the 
radio jets up to distances of $\sim 20$ pc from the core, 
%(i.e., the region where the brightest of the radio 
%knots are likely to be located),
$B$ = 0.1 -- 0.3 G has been 
estimated from independent analyses of the detailed VLBI observations 
(Lobanov 1998; Piner et al.\ 2003). 
Somewhat higher values have been adopted by \cite{bloo} in 
their modeling of the superluminal cores.    
We adopt $B = 0.1$ G as a fiducial value, but consider the plausible range
below.  As noted above, most observations and models give values of
$E_{min,2} \simeq 1$, so we take 1 as its fiducial value. 

The other important physical parameter in Eqs (3) and (4) is the surface 
brightness, $\Sigma$.  A recent data set relevant to this quantity is 
based on 15 GHz Very Long Baseline Arrary (VLBA) imaging of 160 flat-spectrum sources (Kellermann
et al.\ 1998; Zensus et al.\ 2002). These observations, with baselines out 
to 440 million wavelengths (with angular resolution,
$\theta \simeq 0.12$ mas) have shown that 
$\sim$90\% of the sources have an unresolved flux density greater than $0.1$ Jy 
(i.e., $\Sigma \gtrsim 7$), whereas $\sim$45\% of the sources have unresolved 
structures stronger than 0.5 Jy (i.e., $\Sigma \gtrsim 35$) \citep{kova}.
There is presently little evidence in the VLBI samples for a sizable 
population of AGN having $\Sigma > 350$ \citep{kova}. 
This is further corroborated by 
the ongoing space VLBI survey of compact AGN, using the VLBI Space
Observatory Programme  \citep{hira}. A possible explanation for the lack of
 extremely 
high values of $\Sigma$ ($\gtrsim 10^4$) is that such extreme 
objects will form rarely to begin with, and that any such components 
would be very strongly
self-absorbed at typical VLBI frequencies, and therefore relativistic
beaming would be much less effective for flux boosting (see Kellermann 1994).
Thus, we shall adopt here $\Sigma = 10^{1.5 \pm 1}$. Taking $B = 0.1$ G,
and $E_{min,2} = 1$ (see above), we then have from Eq (4), 
$\delta_{\star}$ = $\epsilon (1+z)$, where $\epsilon \simeq$ 2, 4, and 9, 
for $\Sigma$ = 3, 30, and 300, respectively, a range that includes the
vast majority of the bright VLBI knots. % in the vicinity of the nucleus.  

\section {Results and Conclusions}

Evidently, for many radio emission regions close to the nucleus, LEC induced flux 
suppression is expected to set in already at frequencies that are too high for 
SSA to be significant, even if the knots are only modestly superluminal. 
But the same LEC dominance would require larger Doppler factors in the 
case of knots 
having higher surface brightnesses and/or redshifts. 
Fig.\ 1 illustrates the variation 
of $\nu_m$ and $\nu_{SSA}$ across the $\nu$ -- $\delta_0$ plane, for the
plausible ranges 
of $\Sigma$ and $B$. It is interesting to note that for most values of $\delta_0$ 
(i.e., $\delta_0 \gtrsim 4$), $\nu_m$ lies above the typical frequency
range of VLBI observations (5 -- 15 GHz) if $B \gtrsim 0.1$G,
and hence these observations are 
usually made below the spectral turnover due to LEC (although the spectral turnover
due to SSA is yet to set in for most of the knots). For the brighter, albeit 
comparatively rare, knots with $\Sigma \gtrsim 100$, the typical range of 
VLBI frequencies would always lie within the SSA-dominated regime, for 
practically the entire plausible range of $\delta_0$ (Fig.\ 1). In recent 
years, a few VLBI imaging observations have also been possible at higher 
frequencies, e.g., 43--86 GHz (e.g., Krichbaum et al.\ 2001; Jorstad \& Marscher
2003). For such cases, 
the LEC-induced spectral flattening would set in for $\delta_0 \gtrsim 10$, 
except for the brightest VLBI knots ($\Sigma \gtrsim 1000$) which would
nonetheless be synchrotron self-absorbed. Thus, it appears that in typical 
VLBI observations, made at 5 or 15 GHz, most of the emission regions in  
proximity to the nucleus 
(at distances $\la 20$ pc, where the brightest and the fastest moving knots are
 normally 
expected to be found) would appear dimmed either from
flux suppression due to LEC, or alternatively, in the few cases of extreme
brightness, due to SSA. 
As can be seen from Fig.\ 1 and Eq (3), the flux suppression would increase 
rapidly with 
Doppler factor ($\delta$), but its significance would diminish for the 
knots having lower magnetic field and/or higher redshift, or if the value
of $E_{min}$ were smaller.

%\clearpage

\begin{figure}
\plotone{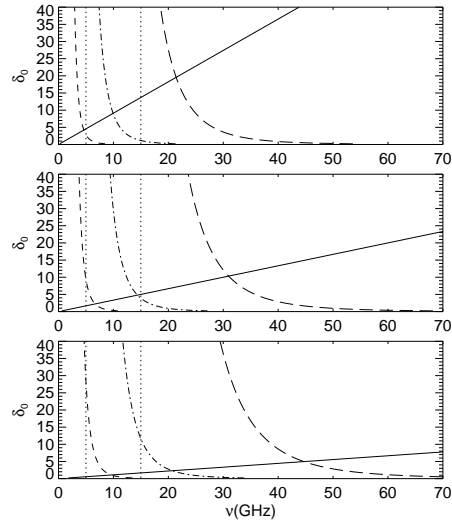}
\caption{Values of $\nu_m$ and $\nu_{SSA}$ against $\delta_0$
and $\Sigma$ for 
$B = 0.03$G ({\it top}), $B = 0.1$G ({\it middle}), $B = 0.3$G ({\it lower});  
$E_{min} = 100$  MeV. 
Solid lines represent $\nu_m$;  
short-dashed, dot-dashed, and long-dashed
curves  represent $\nu_{SSA}$ 
for $\Sigma = 3, 30$, and $300$, respectively.   
LEC dominates whenever the $\nu_m$ line is to the right
of the $\nu_{SSA}$ curve.  The usual VLBI frequencies of 5 and 15 GHz
are shown as dotted lines.
}
\end{figure}

%\clearpage
An observational consequence of the depression of the peak flux density
by a factor $\psi$ (\S 2), arising 
from such (pre-mature) spectral flattening, 
would be a reduction in brightness temperatures, $T_B$, of the VLBI knots, 
compared to the situation in which SSA alone were significant. For the latter 
case, \cite{read} has deduced a realizable limit of $T_{eq} \sim 5 \times 
10^{10}$ K 
(assuming equipartition; also see Singal \& Gopal-Krishna 1985), 
which is about a sixth of the diamagnetic limit 
calculated by \cite{sing}. Other theoretical limits proposed for $T_B$
are $\la 10^{12}$ K, arising from catastrophic inverse Compton cooling 
\citep{kell69} and greater than $2 \times 10^{11}$ K, arising 
from relativistic induced Compton scattering \citep{sinc}. 
Evidence for brightness temperatures being found to  typically be lower 
than these theoretically expected intrinsic brightness temperatures comes 
from the VLBA 2 cm sample of 48 superluminal sources, whose distribution in 
the `Doppler factor -- proper motion' plane is best fitted by a peak intrinsic 
$T_B \simeq 2 \times 10^{10}$ K \citep{cohe}.  This is roughly
half of the equipartition value, and an order of magnitude below
the stricter inverse Compton scattering limit.  The LEC-induced flux suppression
 could possibly
account for the deficit observed vis-{\`a}-vis the equipartition
value.  Note that even though the best-fit value 
of the intrinsic brightness temperatures found by \cite{cohe},
is just a factor of 2 -- 3 below the equipartition temperature limit, it 
makes a huge difference in the energetics.  
The increase in the 
 energies for such  reductions in $T_B / T_{eq}$ is 
$\approx 20 - 200 $ times \citep{read}.%; also see \cite{ting}.
%Lister, Tingay \& Preston 2001; Lister et al.\ 2003).
%The large increase in the 
%required energies for such modest reductions in $T_B / T_{eq}$ is 
%$\approx 20 - 200 $ times \citep{read}.%; also see \cite{ting}.
%Lister, Tingay \& Preston 2001; Lister et al.\ 2003).

To recapitulate, for the typical magnetic field of 0.1 G and usual VLBI
observing 
frequency ($\nu = $ 5 -- 15 GHz), most %the (majority of the) 
radio knots with 
medium-to-low surface brightness ($\Sigma \la 10^2$) would suffer a 
marked suppression of flux density as a result of LEC, if $\delta_0 = \delta/(1+z)$ 
exceeds $\sim$5 and $E_{min,2} \gtrsim 1$. Thus, in such VLBI surveys 
one would expect 
a deficit of knots with high Doppler factors (and, correspondingly, of 
high Lorentz factors, given the rather tight correspondence expected 
between the two, cf.\ Table 1 in Jorstad \& Marscher 2003). This selection 
bias would gradually ease for VLBI surveys made at higher frequencies 
and/or higher $z$ (which corresponds to higher observing 
frequencies in the rest frame of the sources). 

Interestingly, such statistical 
trends may indeed be seen in the recent large VLBI surveys. 
For instance, for a large sample of flat-spectrum sources, the Caltech--Jodrell 
Bank-Flat  (CJF) sample, the measured apparent velocities of 
the VLBI components at 5 GHz show a positive correlation with redshift 
(Fig.\ 1 in Vermeulen et al.\ 2003). While, clearly, this result
does not exclude alternative explanations, it is interesting that there 
is also evidence for a {\it statistical} increase in the apparent velocities 
of superluminal knots with the frequency of the VLBI measurements,
even as the apparent velocity of a given VLBI knot observable at multiple
frequencies would be the same at all of those frequencies. 
As already pointed out by \cite{verm03}, the apparent velocities 
for a CJF sample at 5 GHz are nearly a factor of 2 smaller, on average,
 than those found in 
the 15 GHz VLBA Survey of the same sample \citep{zens03}, and these, 
in turn, are only about half as fast as the velocities found in a VLBI survey 
of 33 blazars at 22 and 43 GHz by \cite{jors}. 
 Their high
frequency survey 
yielded mean apparent velocity of $\sim$11$c$, (taking $H_0 = 65$ km s$^{-1}$ 
Mpc$^{-1}$ and $q_0 = 0.1$). \cite{verm03} suggest that a 
decrease in mean apparent velocity
with decreasing frequency might result from changing SSA turnovers if
these measurements corresponded to an increase in average size or 
distance from the core.  \cite{jors}
 have ascribed the higher velocities for their sample to be an intrinsic 
property of their sources, which are 
$\gamma$-ray emitting blazars. %On the other hand, 
In our scenario, their 
finding abnormally high velocities for that sample may be directly related 
to the 
unusually high observing frequency of their VLBI survey. One observational 
test of our hypothesis would be possible by VLBI monitoring of $\gamma$-ray 
blazars at moderate frequencies, say 5 or 10 GHz; we understand that such
data are likely to become available in the near future, making a detailed 
check possible. 

A crisp prediction of our scenario, which can be tested by quasi-simultaneous
multi-frequency VLBI, is that the knots affected by LEC should have
only a weakly inverted spectrum below $\nu_m$, with a slope of $\sim 1/3$
down to $\nu_{SSA}$. 

The presence of a LEC could also have interesting implications
for the kiloparsec scale jets of quasars, for which bulk relativistic 
motion has been inferred recently from X-ray observations (e.g., 
Tavecchio et al.\ 2000; Celotti, Ghisellini \& Chiaberge 2001). 
 In such extended jets the photon energy density is 
dominated by the cosmic microwave background radiation (CMBR)
 photons, which, after inverse Compton collisions 
with the relativistic jet particles, would be observed at a frequency 
$\nu_{IC} = \delta (1+{\rm cos}\phi^{\prime})\gamma_{min}^2
[1.6\times 10^{11}(1+z)\Gamma]{\rm Hz}$, with $\phi^{\prime}$ the
angle between the jet and the observer in the jet's rest frame
\citep{harr}.

For a typical $z \simeq 1$,  and taking even a fairly modest value of 
$\delta \Gamma \sim 25$, which is close to the expectation value 
for $\Gamma \sim 5$ (e.g., Vermeulen \& Cohen 1994; Jorstad 
\& Marscher 2003), it can be seen that $E_{min} \simeq 100$ MeV 
translates into a spectral turnover of the inverse Compton scattered 
radiation occurring at energies of a few keV. Thus, for many jets detected
with the {\sl Chandra} and {\sl XMM-Newton} telescopes, signatures of the LEC may 
well be present in their X-ray measurements, in the form of
suppressed soft X-ray flux, i.e., a hardened X-ray spectrum.
Moreover, one can expect rapid, correlated variations of 
X-ray flux and spectral slope as $\delta$ changes. 
Clearly, these 
effects would be stronger for extended jets with faster bulk 
motions and/or higher redshifts; a search for these trends would
provide valuable clues. The flux suppression would render the detection 
of extended X-ray jets at higher redshifts more difficult, 
even though the expected surface brightness diminution with $z$ is 
exactly compensated for by the corresponding increase in the CMBR energy 
density (see Schwartz et al.\ 2000).

\acknowledgments{We thank the referee, Alan Marscher, for important
suggestions. %GK would like to 
We thank S.\ Britzen, H.\ Falcke, T.\ Krichbaum,
A.\ Lobanov, S.\ Markoff, A.\ Witzel and A.\ Zensus 
for useful conversations.
%for helpful information about
%the recent VLBI surveys.  PLB thanks H.\ Falcke and S.\ Markoff
%for many useful conversations.  
Work of PLB is mainly supported
through the AUGER theory and membership grant 05 CU1ERA/3 through
DESY/BMBF (Germany). PJW is 
grateful for support from RPE/PEGA at GSU. }
%and 
%for continuing hospitality at Princeton University Observatory.}

%\begin{references}


\begin{thebibliography}{}
\bibitem[\protect\citeauthoryear{Beckert \& Duschl}{1997}]{beck}
Beckert, T., \& Duschl, W.\ J. 1997, A\&A, 328, 95
%\bibitem[\protect\citeauthoryear{Biermann et al.}{2003}]{bier03}
%Biermann, P.\ L., Chirvasa, M., Falcke, H., Markoff, S., 
%\& Zier, Ch. 2003, in Proc.\ 7eme Colloq. Cosmologie, 
%High Energy Astrophysics from and for Space, Paris, 2002 June,
%ed.\ N.\ Sanchez \& H.\ de Vega, in press (astro-ph/0211503)
\bibitem[\protect\citeauthoryear{Biermann et al.}{1995}]{bier95}
Biermann, P.\ L., Strom, R.\ G., \& Falcke, H.  1995, A\&A, 302, 429
\bibitem[\protect\citeauthoryear{Blandford \& McKee}{1976}]{blan}
Blandford, R.\ D., \& McKee, C. 1976, Phys.\ Fluids, 19, 1130
\bibitem[\protect\citeauthoryear{Bloom \& Marscher}{1996}]{bloo}
Bloom, S.\ D., \& Marscher, A.\ P. 1996, ApJ, 461, 657
\bibitem[\protect\citeauthoryear{Britzen et al.}{2001}]{brit}
Britzen, S., Witzel, A., Krichbaum, T.\ P., MacSlow, T.,
\&  Matveyenko, L.\ I. 2001, AstrL, 27, 1
\bibitem[\protect\citeauthoryear{Carilli et al.}{1991}]{cari}
Carilli, C.\ L.,  Perley, R.\ A., Dreher, J.\ W., \& Leahy, P.
1991, ApJ, 383, 554
\bibitem[\protect\citeauthoryear{Celotti \& Fabian}{1993}]{celo93}Celotti, A., \& Fabian, A.\ C. 1993, MNRAS, 264, 22
\bibitem[\protect\citeauthoryear{Celotti et al.}{2001}]{celo01}
Celotti, A., Ghisellini, G., \& Chiaberge, M. 2001, MNRAS, 321, L1
%\bibitem[\protect\citeauthoryear{Chirvasa}{2002}]{chir}
%Chirvasa, M. 2002, M.S.\ thesis, Univ.\ Bucharest
\bibitem[\protect\citeauthoryear{Cohen et al.}{2003}]{cohe}
Cohen, M.\ H., et al. 2003, in ASP Conf.\ Ser.\ 300, Radio Astronomy 
at the Fringe, ed.\ J.\ A. Zensus, M.\ H.\ Cohen, \& E.\ Ros (San Francisco: ASP), 177
\bibitem[\protect\citeauthoryear{Donea \& Biermann}{1996}]{done}
Donea, A.\ C., \& Biermann, P.\ L. 1996, A\&A, 316, 43
\bibitem[\protect\citeauthoryear{Duschl \&  Lesch}{1994}]{dusc}
Duschl, W.\ J., \& Lesch, H. 1994, A\&A, 286, 431
\bibitem[\protect\citeauthoryear{Falcke et al.}{1995}]{falc95}
Falcke, H., Malkan, M.\ A., \& Biermann, P.\ L. 1995, A\&A, 298, 375 
\bibitem[\protect\citeauthoryear{Falcke et al.}{1993}]{falc93}
Falcke, H., Mannheim, K., \& Biermann, P.\ L. 1993, A\&A, 278, L1 
\bibitem[\protect\citeauthoryear{Ghisellini et al.}{1992}]{ghis}
Ghisellini, G., Celotti, A., George, I.\ M., \& Fabian, A.\ C. 1992, MNRAS, 258, 776
\bibitem[\protect\citeauthoryear{Ginzburg \& Syrovatskii}{1969}]{ginz}
Ginzburg, V.\ L., \& Syrovatskii, S.\ I. 1969, ARAA, 7, 375
\bibitem[\protect\citeauthoryear{Hardcastle}{2001}]{hard}
Hardcastle, M.\ J. 2001, A\&A, 373, 881
%\bibitem[\protect\citeauthoryear{{}Harris, D.\ E. 2003, in Radio Astronomy at the Fringe,
%eds.\ J.\ A. Zensus, M.\ H.\ Cohen \& E.\ Ros, ASP Conf Ser., Vol.\ 300,
%(San Francisco: ASP), 151
\bibitem[\protect\citeauthoryear{Harris \& Krawczynski}{2002}]{harr}
Harris, D.\ E., \& Krawczynski, H. 2002, ApJ, 565, 244
\bibitem[\protect\citeauthoryear{Hirabayashi et al.}{2000}]{hira}
Hirabayashi, H., et al. 2000, PASJ, 52, 997
\bibitem[\protect\citeauthoryear{Jones \& O'Dell}{1977}]{jone}
Jones, T.\ W., \& O'Dell, S.\ L. 1977, A\&A, 61, 291
\bibitem[\protect\citeauthoryear{Jorstad \& Marscher}{2003}]{jors}
Jorstad, S., \& Marscher, A.\ P. 2003, in ASP Conf.\ Ser.\ 300, Radio Astronomy 
at the Fringe, ed.\ J.\ A. Zensus, M.\ H.\ Cohen, \& E.\ Ros (San Francisco: ASP),  89
\bibitem[\protect\citeauthoryear{Kellermann}{1994}]{kell94}
Kellermann, K.\ I. 1994,  in Compact Extragalactic Radio Sources,
eds.\ J.\ A.\ Zensus \& K.\ I.\ Kellermann (Green Bank: NRAO), 245 
\bibitem[\protect\citeauthoryear{Kellermann}{2003}]{kell03}
Kellermann, K.\ I. 2003,  in ASP Conf.\ Ser.\ 300, Radio Astronomy 
at the Fringe, ed.\ J.\ A. Zensus, M.\ H.\ Cohen, \& E.\ Ros (San Francisco: ASP), 185
\bibitem[\protect\citeauthoryear{Kellermann \& Pauliny-Toth}{1969}]{kell69}
Kellermann, K.\ I., \& Pauliny-Toth, I.\ I.\ K. 1969, ApJ, 155, L71
\bibitem[\protect\citeauthoryear{Kellermann et al.}{1998}]{kell98}
Kellermann, K.\ I., Vermeulen, R.\ C.,
 Zensus, J.\ A., \& Cohen, M. H. 1998, AJ, 115, 1295
\bibitem[\protect\citeauthoryear{Kovalev}{2003}]{kova}Kovalev, Y.\ Y. 2003,  in ASP Conf.\ Ser.\ 300, Radio Astronomy 
at the Fringe, ed.\ J.\ A. Zensus, M.\ H.\ Cohen, \& E.\ Ros (San Francisco: ASP), 65
\bibitem[\protect\citeauthoryear{Krichbaum et al.}{2001}]{kric
}Krichbaum, T.\ P., Graham, D.\ A., Witzel, A., Zensus, J.\ A., 
Greve, A., Grewing, M., Marscher, A., \& Beasley, A.\ J. 2001, in ASP Conf.\ Ser.\ 250,
Particles and Fields in Radio Galaxies, ed.\ R.\ A.\ Laing \& K.\ M.\ Blundell
 (San Francisco: ASP), 184
\bibitem[\protect\citeauthoryear{Leahy et al.}{1989}]{leah}
Leahy, J.\ P., Muxlow, T.\ W.\ B., \& Stephens, P.\ W. 1989, MNRAS, 239, 401
%\bibitem[\protect\citeauthoryear{{}Lister, M.\ L., Tingay, S.\ J., Preston, R.\ A. 2001, ApJ, 554, 964 
%\bibitem[\protect\citeauthoryear{{} Lister, M.\ L., Kellermann, K.\ I., Vermeulen, R.\ C., Cohen, M.\ H.,
% Zensus, J.\ A., \& Ros, E.  2003, ApJ, 584, 135
\bibitem[\protect\citeauthoryear{Lobanov}{1998}]{loba}Lobanov, A.\ P. 1998, A\&A, 330, 79
\bibitem[\protect\citeauthoryear{Lovelace}{1976}]{love}Lovelace, R.\ V.\ E. 1976, Nature, 262, 649
%\bibitem[\protect\citeauthoryear{Mahadevan}{1998}]{maha}Mahadevan, R. 1998, Nature, 394, 651
\bibitem[\protect\citeauthoryear{Marscher}{1987}]{mars}Marscher, A.\ P. 1987, in Superluminal Radio Sources, ed.\ J.\ A.\ Zensus \& 
T.\ J.\ Pearson (Cambridge: Cambridge UP), 280
\bibitem[\protect\citeauthoryear{Matveyenko}{1980}]{matv}
Matveyenko, L.\ I. et al. 1980, SvAL, 6, 42
\bibitem[\protect\citeauthoryear{Pacholczyk}{1970}]{pach}Pacholczyk, A.\ G. 1970, Radio Astrophysics 
(San Francisco: Freeman)
\bibitem[\protect\citeauthoryear{Piner et al.}{2003}]{pine}
Piner, B.\ G., Unwin, S.\ C., Wehrle, A.\ E., Zook, A.\ C.,
Urry, C.\ M., \& Gilmore, D.\ M. 2003, ApJ, 588, 716
\bibitem[\protect\citeauthoryear{Readhead}{1994}]{read}Readhead, A.\ C.\ S. 1994, ApJ, 426, 51
%\bibitem[\protect\citeauthoryear{Salter \& Brown}{1988}]{salt}
%Salter, C.\ J., \& Brown, R.\ L. 1988, in Galactic and Extragalactic
%Radio Astronomy, ed.\ G.\ L.\ Verschuur \& K.\ I.\ Kellermann
%(2d ed., New York: Springer), 1
\bibitem[\protect\citeauthoryear{Schwartz  et al.}{2000}]{schw}
Schwartz, D.\ A., et al.\ 2000, ApJ, 540, L69
\bibitem[\protect\citeauthoryear{Sincell \& Krolik}{1994}]{sinc}
Sincell, M.\ W., \& Krolik, J.\ H. 1994, ApJ, 430, 550
\bibitem[\protect\citeauthoryear{Singal}{1988}]{sing}Singal, A.\ K. 1988, A\&A, 155, 242
\bibitem[\protect\citeauthoryear{Singal \& Gopal-Krishna}{1985}]{sigk}
Singal, A.\ K., \& Gopal-Krishna 1985, MNRAS, 215, 383
\bibitem[\protect\citeauthoryear{Tavecchio et al.}{2000}]{tave}
Tavecchio, F., Maraschi, L, Sambruna, R.\ M., \& Urry, C.\ M. 2000, ApJ, 544, L23
%\bibitem[\protect\citeauthoryear{Tingay  et al.}{2001}]{ting}
%Tingay, S.\ J., et al.\ 2001, ApJ, 549, L55
\bibitem[\protect\citeauthoryear{Vermeulen \& Cohen}{1994}]{verm94}
Vermeulen, R.\ C., \& Cohen, M.\ H. 1994, ApJ, 430, 467
\bibitem[\protect\citeauthoryear{Vermeulen et al.}{2003}]{verm03}
Vermeulen, R.\ C., Britzen, S., Taylor, G.\ B., Pearson, T.\ J.,
Readhead, A.\ C.\ S., Wilkinson, P.\ N., \& Browne, I.\ W.\ A. 2003, 
 in ASP Conf.\ Ser.\ 300, Radio Astronomy 
at the Fringe, ed.\ J.\ A. Zensus, M.\ H.\ Cohen, \& E.\ Ros (San Francisco: ASP), 43
\bibitem[\protect\citeauthoryear{Wardle}{1977}]{ward}Wardle, J.\ F.\ C. 1977, Nature, 269, 563
\bibitem[\protect\citeauthoryear{Zensus et al.}{2002}]{zens02}
Zensus, J.\ A., Ros, E., Kellermann, K.\ I.,
 Cohen, M.\ H., Vermeulen, R.\ C., \& Kadler, M. 2002, AJ, 124, 662
\bibitem[\protect\citeauthoryear{Zensus et al.}{2003}]{zens03}
Zensus, J.\ A.,  et al. 2003,  in ASP Conf.\ Ser.\ 300, Radio Astronomy 
at the Fringe, ed.\ J.\ A. Zensus, M.\ H.\ Cohen, \& E.\ Ros (San Francisco: ASP),  27
%\end{references}

\end{thebibliography}
\end{document}